\documentstyle[twocolumn,aps]{revtex}
 \input psfig
 \newdimen\figwidth \figwidth=8.5cm   

\begin{document}

 \draft


 \title{A Stochastic Diffusion Model of Climate Change} 
 \author{Jon D. Pelletier}
 \address{Department of Geological Sciences, Snee Hall, Cornell
University, Ithaca, NY 14853}
 \maketitle
 \vspace{1cm}

 \narrowtext

 \section{Introduction} 
There is widespread agreement that the obliquity and precession
of the Earth's
orbit result in a significant portion of
global temperature variance at frequencies near
40,000 and 20,000 years.
Aside from this ``external'' forcing of the climate system,
there is no concensus on the physical processes governing the remainder
of the variance.
It is generally agreed that the variance resulting from orbital forcing
is superimposed on a continuous ``background''
spectrum with a variance that increases with
decreasing frequency (Kerr 1978).  
Most
attempts to model this background spectrum assume a nonlinear response of
the climate system to the orbital forcing (Bhattacharya, Ghil, and Vulis
1982; Imbrie and Imbrie 1980; Le Treut and Ghil
1983; Nicolis 1993).
Each of these studies has succeeded in reproducing only a
limited portion of the spectrum of climatic variations.

Several studies have recognized that the low-frequency
portion of the power spectrum, $S(f)$,
of temperature variations
inferred from ice and ocean cores has the form of a Lorentzian distribution:
(Hasselmann 1971; Komintz and Pisias
1979; Lovejoy and Shertzer
1986)
\begin{equation}
S(f)\propto \frac{1}{f^{2}+f_{0}^{2}}
\end{equation}
This function is constant for very low
frequencies and proportional to
$f^{-2}$ for $f \gg f_{0}$.
Hasselmann (1971)
and Komintz and Pisias (1979) have
suggested that a stochastic
model may be the most appropriate model for this spectrum
since a Lorentzian spectrum can result from dynamics in which
a macroscopic variable is the sum of
uncorrelated random pulses with a negative feedback
mechanism limiting the variance at low frequencies.

At shorter time scales it is recognized that there is ``persistence'' or
correlations in time in meteorological time series over a range of time
scales. 
Persistence means that warm years (or weeks or
months) are, more often 
than not, followed by warm years and cold years by cold years.
Hurst, Black, and Simaika (1965)
presented studies of these correlations using the rescaled-range technique
(see Feder 1991 for an introduction).
Hurst found that
time series of annual mean temperature produced a power-law 
rescaled-range plot with an average exponent of 0.73. 
No persistence would yield an exponent of 0.5.
Numerical studies
have shown that series with a Hurst
exponent of 0.73
have a power spectrum proportional to
$f^{-\frac{1}{2}}$ (Higuchi 1990; Gomes da Silva and Turcotte 1994; Malamud
and Turcotte
1995).

In this paper we present a model that provides a specific 
physical mechanism for the entire background spectrum from time scales
of 1 day to 1 million years. We present spectral analyses of paleoclimatic
proxy data and instrumental data that support the 
observation of a low-frequency
Lorentzian spectrum and a higher-frequency spectrum proportional to
$f^{-\frac{1}{2}}$. At very high frequencies we found a distinct difference
between the power spectra of continental and maritime stations. 
Continental stations exhibit power spectra proportional to 
$f^{-\frac{3}{2}}$ at time scales less than one month while 
the spectra of maritime
stations remain proportional to $f^{-\frac{1}{2}}$ down to time scales of
one day. Our model is based upon an analytic approach to modeling the
stochastic diffusion of heat in the atmosphere and ocean.
The difference between continental and maritime stations arises because
the air mass above maritime stations exchanges heat with both the 
atmosphere above and the ocean below while the air mass above continental
stations exchanges heat with only the atmosphere above it.

``If turbulent transfer in a system is dominated by eddies much smaller 
than the system 
size, random convective action of turbulent eddies will be analogous
to the molecular agitation responsible for molecular diffusion''
(Moffatt 1983). In this approximation, turbulent transfer can be modeled
as a stochastic diffusion process (Csanady 1980).
This is equivalent to the
Lagrangian theory of turbulence for times long compared to the Lagrangian
time scale (time below which particle velocities are autocorrelated)
(Tennekes and Lumley 1972; Gifford 1982). In models of climate
change, it is common to model
the turbulent transfer of heat as a deterministic diffusion process
(Ghil 1983).
The stochasticity
of turbulent transfer results in temperature fluctuations from equilibrium
not present in a deterministic model of turbulent heat transfer. We
will present the power spectrum of temperature fluctuations from 
equlibrium resulting from stochastic heat transport
in a two-layer geometry appropriate to the atmosphere and
ocean. 

The model we present was first solved by van
Vliet, van der Ziel, and Schmidt (1980) to determine 
the power spectrum of variations due to the
stochastic diffusion of heat in a metallic
film in thermal equilibrium with a substrate. Temperature variations in the
film and substrate occur as a result of fluctuations in the heat transport 
by 
electrons undergoing Brownian motion. The top of the film absorbs and emits
blackbody radiation. In this paper we use van Vliet et al.'s model exactly
as they presented it with the atmosphere as the metallic film and
the ocean as the substrate. Turbulent eddies in the atmosphere and ocean
are analagous to the electrons undergoing Brownian motion in a metallic
film in contact with a substrate. 
The model studied by van Vliet et al. (1980), with
physical constants appropriate to the atmosphere and ocean,
yield a power spectrum in agreement with that of climatic variations
recorded in instrumental records
and inferred from ocean and ice cores from time scales
of one day to one million years.

\section{Observations of Climatic Variations}

 \begin{figure}
 \psfig{figure=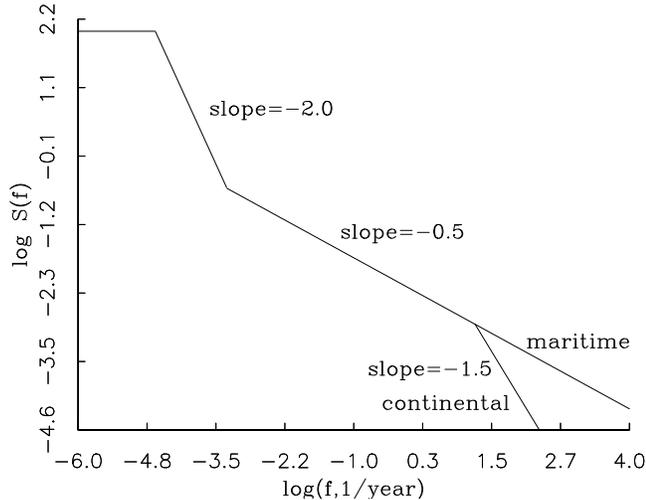,width=\figwidth}
 \caption{Logarithm of the power spectrum of atmospheric temperature
variance predicted by the model as a function of
the logarithm of the
frequency in years$^{-1}$. The crossover frequencies observed in the
climatological record are, from left to right,
$f_{0}=\frac{1}{40,000 years}, f_{1}=\frac{1}{2,000 years},
f_{2}=\frac{1}{1 month}$.}
 \label{one}
 \end{figure}

In Figure~\ref{one} 
we present of the logarithm of the
power spectrum of the variations in
temperature as a function of the logarithm of the frequency
predicted by the model and supported by observational data.
At low frequencies the power spectum is constant. Above
$f\approx\frac{1}{40,000\ years}$ the
power spectrum is proportional to $f^{-2}$.
Above $f\approx\frac{1}{2,000\ years}$ the
power spectrum is proportional to $f^{-\frac{1}{2}}$.
At very high frequencies
(above $f\approx\frac{1}{1\ month}$) the spectrum varies as
$f^{-\frac{3}{2}}$ for continental stations and remains proportional to
$f^{-\frac{1}{2}}$ for maritime stations.
The crossover frequencies quoted above are those observed in
the climatological record. The model predicts crossover frequencies which
are close to those observed in the climatological record (two out of three
are within a factor of two of the observed frequencies).
All of the power spectra that we present in this paper are plotted after
taking the logarithm of the power spectrum and of the frequency against which
the spectrum is plotted. All power-law functions appear as
straight lines with a slope equal to the exponent of the power-law.
The frequency unit of all observational data is years$^{-1}$.

 \begin{figure}
 \psfig{figure=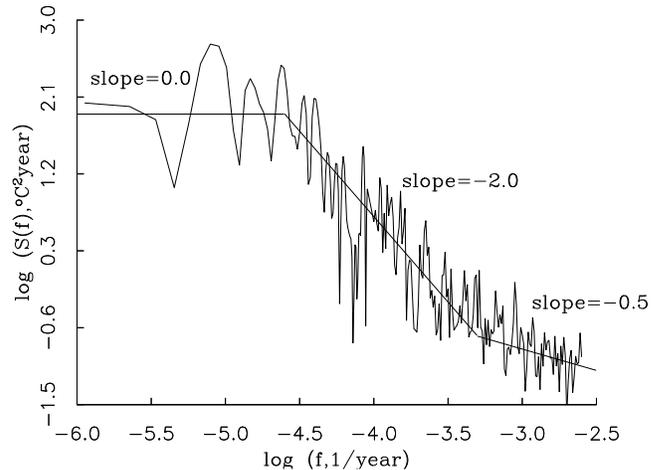,width=\figwidth}
 \caption{Logarithm of the normalized
Lomb periodogram of the temperature inferred from
the Vostok ice core as a
function of the logarithm of the frequency in years$^{-1}$}
 \label{two}
 \end{figure}

Figure~\ref{two} shows the logarithm of the normalized Lomb periodogram of the
Vostok $\delta D$ record extending back 220,000 years
converted into degrees Celsius by the conversion factor 5.6\% per degree
Celsius (Jouzel et al. 1987). We obtained the spectrum from M. Ghil and
P. Yiou (1994). 
It is not possible to directly estimate the power spectrum of the
Vostok record with the FFT
since the data are unevenly sampled. Numerical Recipes
suggests the use of the Lomb Periodogram for such data (Press et al. 1992).
The periodogram
shows a constant low-frequency region that changes to a
region proportional to $f^{-2}$ above $f\approx\frac{1}{40,000\ years}$.
Above
$f\approx\frac{1}{2,000\ years}$ the power spectrum of temperature variations
is
$\propto f^{-\frac{1}{2}}$. To reduce the scatter of the periodogram
at high frequencies, we averaged the periodogram in logarithmically-spaced
bins of size $\log f=0.01$ above $\log f=-4.0$.

 \begin{figure}
 \psfig{figure=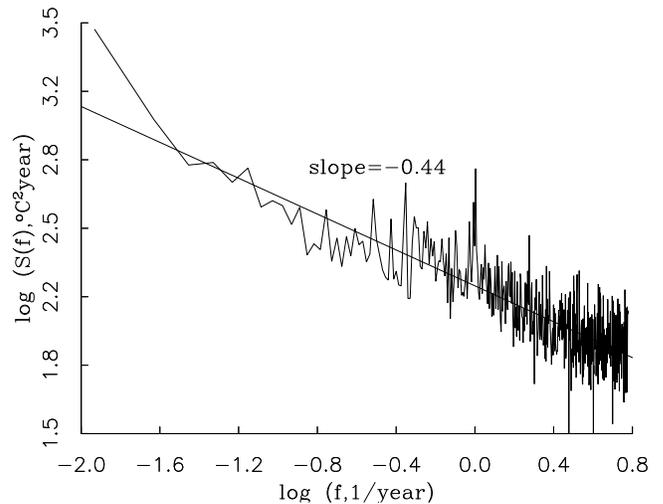,width=\figwidth}
 \caption{The average unnormalized power spectrum
of 94 monthly temperature time series on a log-log plot as a function of
frequency in years$^{-1}$.
}
 \label{three}
 \end{figure}

In Figure~\ref{three}
we present the logarithm of the unnormalized
power spectra of temperature time series
as a function of the logarithm of the frequency taken instrumentally
at higher frequencies. 
We plot the logarithm of the
average power spectrum
of time series of monthly mean temperature from 94 stations worldwide
with the yearly trend removed.
The power spectra were computed by taking the modulus squared of the
complex Fourier coefficients obtained by computing
the Fast Fourier Transform using the Numerical Recipes
routine realft (Press et al. 1992).
We computed the power spectra of all
complete temperature series of length greater than or equal to 1024 months
from the
climatological database compiled by Vose et al. (1992).
The yearly trend was removed by subtracting from each monthly
data point the average temperature for that month in the 86 year record for
each station. All of the power spectra were then averaged at equal
frequency values.
The data yield a straight-line
least-square fit
with slope close to -0.5 indicating that
$S(f)\propto f^{-\frac{1}{2}}$ in this frequency region.
This is consistent with the results of Hurst, Black, and Simaika (1965)
who analyzed correlations at instrumental time scales with the 
rescaled-range method.
Numerical studies have shown that a time series which yields a power-law 
rescaled-range plot with Hurst exponent $H=0.73$ has a power spectrum
$S(f)\propto f^{-\frac{1}{2}}$
(Higuchi 1990; Gomes da Silva and Turcotte 1994; Malamud
and Turcotte
1995). $f^{-\frac{1}{2}}$ power spectra have also been observed in variations
of atmospheric humidity on time scales of days to years (Vattay and Harnos
1994) and in tree-ring widths (a proxy for precipitation)
up to several thousands of years (Pelletier 1995). 

 \begin{figure}
 \psfig{figure=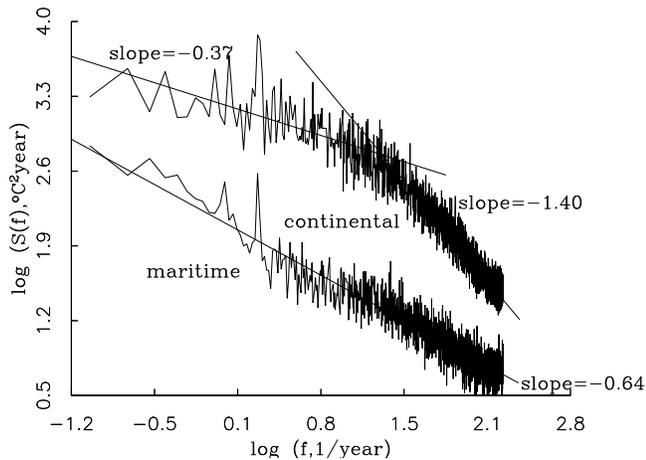,width=\figwidth}
 \caption{Averaged unnormalized
power spectrum of 50 continental and 50 maritime daily temperature
time series on a log-log plot
as a function of frequency in
years$^{-1}$. The crossover frequency for the continental specta
is $f_{2}=\frac{1}{1 month}$.
}
 \label{four}
 \end{figure}

In Figure~\ref{four}
we present the average unnormalized
power spectrum of time series of
daily mean temperature (estimated by taking the average of the maximum and
minimum temperature of each day)
from 50 continental and 50 maritime stations
over 4096 days estimated by computing the Fast Fourier
transform as before.
Maritime stations were sites on small islands far from any
large land masses. Continental stations, conversely, were well inland on
large continents, far from any large bodies of water.
We chose 50 stations at random from the
complete records (those with
greater than 4096 nearly consecutive days of data) provided
by the Global Daily Summary database compiled by the National Climatic Data
Center (1994). We found no records in the database that were without
at least a few weeks worth of missing days over the 14 years of data covered
by the longest of the station records. We filled in the missing days with
the same value as the previous day. This will not introduce any discernible
error since the erroneous datapoints make up at most a fraction of one
percent of each series.
The yearly periodicity in these
data were removed by subtracting from each station's temperature time
series a converged least-squares fit (using Numerical Recipes
(Press et al. 1992)
routine mrqmin)
to a function of the form
\begin{equation}
T_{i}=T_{av}+A\ \cos(\frac{2\pi}{365}i+\phi)
\end{equation}
where $i$ is the number of the day in the year and $T_{av}$, $A$, and $\phi$
were the fitting parameters. This procedure is a standard one for
subtracting the yearly periodicity in a meteorological time series
(Janosi and Vattay 1992).
Continental stations exhibit a $f^{-\frac{3}{2}}$ high-frequency region.
Maritime stations exhibit $f^{-\frac{1}{2}}$ scaling up to the
highest frequency.

\section{Turbulent Transport as a Stochastic Diffusion Process}
In most models of climate change, turbulent transfer of heat energy in
the atmosphere and ocean is modeled by a diffusion process
(Ghil 1983). This assumption
is supported by the self-consistent results of
experimental studies of tracer dispersion which
parameterize transport as a diffusion process. The resulting measurements
of eddy diffusivity are nearly constant as a function of depth in the ocean
and height in the atmosphere. If heat transport was dominated by
convection
currents of the same scale as the height of the atmosphere
or the depth of the ocean,
the resulting diffusion coefficient from such large-scale tracer
studies would be greatly inhomogeneous. Studies of the dispersion
of Tritium vertically in the ocean have resulted in a depth-independent
(except for the mixed layer extending down to about 100\ m) eddy diffusivity
of about $6$x$10^{-6}$ m$^{2}$/s
(Garrett 1984).
Vertical diffusivity in the atmosphere is
considerably more uncertain, but a height-independent
order of magnitude estimate of
1 m$^2$/s for stable air conditions has been quoted by a couple of studies
(Pleune 1990; Seinfeld 1986). 
Radar studies of trace gases in
the middle atmosphere obtained the same diffusivity to
within an order of magnitude from 5-10\ km (Fukao et al. 1994).
Since the time scale of horizontal diffusion in the atmosphere and ocean
is so much smaller
than the time scale of vertical diffusion, diffusion of heat into
and out of a local air mass is one-dimensional. For this reason,
we consider only the variations in local temperature resulting from
heat exchange vertically in the atmosphere and ocean. 

Although coherent air motions
exist which lead to large-scale periodic motions of the atmosphere, these
motions are strongest horizontally and are predominant above the troposphere 
where less than half of the heat capacity of the atmosphere
resides (Dunkerton 1993). 
We will assume that such oscillatory air motions transport at most a small
fraction of the total heat transported vertically in the atmosphere and ocean.
If most of the heat is transported by eddies much smaller than the 
height of the atmosphere and the depth of the ocean, turbulent transport
can be modeled as a diffusion process. 

\section{Fluctuation Theory of Climate Change}
A stochastic diffusion process can be studied analytically by adding a noise
term to the flux of a deterministic diffusion equation
(van Kampen 1981): 
\begin{equation}
\rho c \frac{\partial \Delta T}{\partial t}=-\frac{\partial J}{\partial x}
\label{eqthree}
\end{equation}
\begin{equation}
J=-\sigma\frac{\partial \Delta T}{\partial x} + \eta(x,t)
\label{hi}
\end{equation}
where $\Delta T$ is the fluctuation in temperature from equilibrium and
the mean and variance of the noise is given by
\begin{equation}
\langle \eta(x,t)\rangle=0
\end{equation}
\begin{equation}
\langle \eta(x,t)\eta(x',t')\rangle\propto\sigma(x)\langle T(x)
\rangle^{2}\delta(x-x')\delta(t-t')
\label{eqsix}
\end{equation}

 \begin{figure}
 \psfig{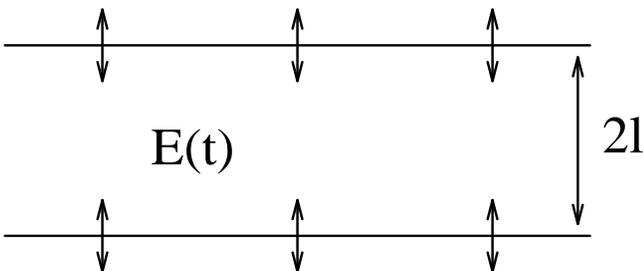}
 \caption{Geometry of the diffusion calculation detailed in the text.}
 \label{oneddiff}
 \end{figure}

We will calculate the power spectrum of temperature fluctuations
in a layer of width 
$2l$ of an infinite, one-dimensional, homogeneous space. The
presentation we give is similar to that of Voss and Clarke (1976).
The variations in total heat energy
in the layer of width 
$2l$ is determined by the heat flow across the boundaries. 
Figure~\ref{oneddiff} illustrates the geometry of the layer exchanging
thermal energy with diffusing regions above and below it.   
A diffusion process has a
frequency-dependent correlation length
$\lambda=(\frac{2D}{f})^{\frac{1}{2}}$ (Voss and Clarke 1976). 
Two different situations arise as a consequence of the length scale, 2l,
of the geometry. For very high frequencies, $\lambda <<2l$. In that case,
the fluctuations in heat flow across the two boundaries are independent.
For low frequencies, $\lambda >>2l$ and the boundaries fluctuate coherently.
First we consider high frequencies. Since the boundaries fluctuate
independently, we can consider the flow across one boundary only.
The flux of heat energy is given by eq. (\ref{hi}).
Its Fourier transform is given by
\begin{equation}
J(k,\omega)=\frac{i\omega\eta(k,\omega)}{\alpha k^{2}-i\omega}
\end{equation}
where $\alpha =\frac{\sigma}{\rho c}$ is the vertical thermal diffusivity.
The flux of heat energy out of the layer at the boundary at $x=l$ (the
other boundary is located at $x=-l$)
is the rate of change of the
total energy in the layer
$E(t)$: $\frac{dE(t)}{dt}=J(l,t)$. The Fourier transform of
this equation is
\begin{equation}
E(\omega)=-\frac{i}{(2\pi)^{\frac{1}{2}}\omega}\int_{-\infty}^{\infty}
dk\ e^{ikl}J(k,\omega)
\end{equation}
Therefore, the power spectrum of variations in $E(t)$,
$S_{E}(\omega)=<|E(\omega)|^{2}>$, is
\begin{equation}
S_{E}(\omega)\propto\int_{-\infty}^{\infty}\frac{dk}{D^{2}k^{4}+\omega^{2}}
\propto\omega^{-\frac{3}{2}}
\end{equation}
Since $\Delta T\propto\Delta E$, $S_{T}(\omega)\propto\omega^{-\frac{3}{2}}$
also.

If we include the heat flux out of both boundaries, the rate of change of
energy in the layer will be given by the difference in heat flux:
$\frac{dE(t)}{dt}=J(l,t)-J(-l,t)$. The Fourier transform of
$E(t)$ is now
\begin{equation}
E(\omega)=\frac{1}{(2\pi)^{\frac{1}{2}}\omega}\int_{-\infty}^{\infty}
dk\ sin(kl)J(k,\omega)
\end{equation}
Then,
\begin{eqnarray}
S_{T}(\omega)\propto
S_{E}(\omega)\propto\int_{-\infty}^{\infty}\frac{dk\ sin^{2}(kl)}
{D^{2}k^{4}+\omega^{2}} 
\nonumber\\ \propto\omega^{-\frac{3}{2}}(1-e^{-\theta}(\sin\theta
+\cos\theta))
\end{eqnarray}
where $\theta =(\frac{\omega}{\omega_{o}})^{\frac{1}{2}}$ and
$\omega_{o}=D/2l^{2}$ is the frequency where the correlation length is equal
to the width of the layer.
When $\lambda <<2l$, the above expression reduces to $S_{T}(\omega)\propto
\omega^{-\frac{3}{2}}$. When $\lambda >>2l$, $S_{T}(\omega)\propto
\omega^{-\frac{1}{2}}$ (Voss and Clarke 1976).

In Section 2 we presented evidence that
continental stations exhibit a $f^{-\frac{3}{2}}$ high-frequency region
and maritime stations exhibit $f^{-\frac{1}{2}}$ scaling up to the
highest frequency. This observation can be interpreted in terms of the
diffusion model presented above. The power spectrum of temperature
variations in an air mass exchanging heat by one-dimensional
stochastic diffusion is 
$\propto f^{-\frac{1}{2}}$ if the air mass is bounded by two diffusing regions
and is $\propto f^{-\frac{3}{2}}$ if it interacts
with only one. The maritime stations
have a $f^{-\frac{1}{2}}$ power spectrum up to the highest frequency
because the air mass above a maritime station exchanges heat with both
the atmosphere above and the ocean below. The fluctuation
calculation appropriate for maritime stations is one in which the coherent
fluctuations from two boundaries are considered as in the calculation
of the $f^{-\frac{1}{2}}$ spectrum.
The air mass above continental
stations exchanges heat energy only with the atmosphere
above it. The calculation appropriate for continental stations is the 
one-boundary
model which predicts the observed $f^{-\frac{3}{2}}$ spectrum.
At low frequencies, horizontal diffusive heat exchange between continental
and maritime air masses limits the variance
of the continental stations. Low-frequency fluctuations of continental
stations may correspond to fluctuations in the average temperature of
the atmosphere above the oceans, giving a $f^{-\frac{1}{2}}$ spectrum.

At lower frequencies, the atmosphere and ocean achieve thermal
equilibrium. The variance in temperature of the atmosphere and ocean
is then determined by the radiation boundary condition.
The fluctuating temperature of 
the atmosphere adds and subtracts heat from the 
atmosphere and ocean through a linear radiation boundary condition; the heat
flux out of the atmosphere is proportional to the temperature of the 
atmosphere. This results in temperature and irradiance variations
with a random walk
($f^{-2}$) spectrum.

The fluctuating input and output of heat in the $f^{-2}$ region will
cause large variations from equilibrium. When the temperature of the
atmosphere and ocean becomes larger than the equilibrium temperature, it will
radiate, on average, more heat than at equilibrium. Conversely, when the
temperature of the atmosphere and ocean wanders lower than the equilibrium
temperature, less heat is radiated. This negative feedback
limits the variance at low frequencies resulting in a constant power spectrum
at very low frequencies.

\begin{figure}
 \psfig{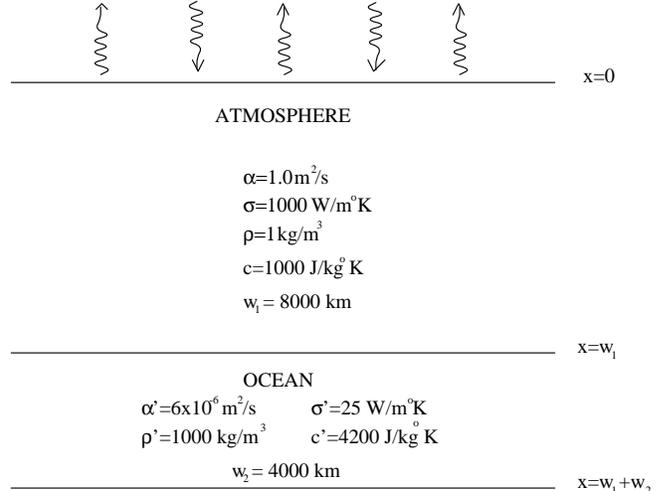}
 \caption{Geometry and constants of the model described in the text.}
 \label{five}
 \end{figure}

To show this, we consider a coupled atmosphere-ocean 
model with an atmosphere of uniform density (equal to the
density at sea level) in contact with an ocean of
uniform density. The height of our model atmosphere is the scale
height of the atmosphere
(height at which the pressure falls by a factor of $e$ from its
value at sea level). Figure~\ref{five} illustrates the geometry
and constants chosen (where $\sigma$ is the vertical heat conductivity, $\rho$
is the density, $c$ is the specific heat per
unit mass, $\alpha$ is the vertical
thermal diffusivity, and $g$ is the thermal conductance of heat out of
the Earth by emission
of radiation). Primed constants denote values for the ocean.
The physical constants which enter the model are the thermal conductance by
emission of radiation and the density, specific heat,
vertical thermal diffusivity,
and depth of the ocean and atmosphere. The density and
specific heat of air and water are well-known constants. We chose an ocean
depth of 4\ km and an atmospheric height equal to the scale height of 8\ km as
used by Hoffert, Callegari, and Hsieh (1980) in their climate modeling studies.
The turbulent transfer of heat vertically in the ocean by diffusion
and advection is commonly
parametrized in studies of climate change as a diffusion process
with an effective thermal diffusivity as we assume in this model (Ghil 1983).
The diffusion coefficients of the atmosphere and ocean we
employ were discussed in
Section 3.

The equation for temperature fluctuations in space and time in the model
is eq.(\ref{eqthree}-\ref{eqsix}):
\begin{equation}
\frac{\partial \Delta T(x,t)}{\partial t}-\alpha (x)
\frac{\partial ^{2} \Delta T(x,t)}
{\partial x^{2}}
=-\frac{\partial\eta(x,t)}{\partial x}
\end{equation}
with
\begin{equation}
\langle \eta(x,t)\rangle=0
\end{equation}
\begin{equation}
\langle \eta(x,t)\eta(x',t')\rangle\propto\sigma(x)\langle T(x)
\rangle^{2}\delta(x-x')\delta(t-t')
\end{equation}

The boundary conditions are that there be no heat flow out of
bottom of the ocean and continuity of temperature
and heat flux at the atmosphere-ocean boundary:
\begin{equation}
\sigma '
\frac{\partial T}{\partial x}|_{x=w_{2}} = 0
\end{equation}
\begin{equation}
\Delta T(x=w_{1}^{+})=\Delta T(x=w_{1}^{-})
\end{equation}
\begin{equation}
\sigma \frac{\partial \Delta T}{\partial x}|_{x=w_{1}^{-}}=
\sigma ' \frac{\partial \Delta T}{\partial x}|_{x=w_{1}^{+}}
\end{equation}

At the top of the atmosphere we will impose a blackbody radiation boundary
condition.
Most $(65\%)$ of the energy incident on the Earth is emitted as long-wavelength
blackbody radiation from the H$_{2}$O and CO$_{2}$ in the atmosphere
(Peixoto and Oort 1992). This
heat emitted from the atmosphere is dependent on the temperature of the
atmosphere at the point of emission according to the Stefan-Boltzmann
law. It is common practice to assume that temperature
variations from equilibrium are small (the global mean temperature has
fluctuated by only about eight degrees Celcius during the last glaciation).
Within a linear approximation, the emitted temperature will be
proportional to the temperature difference from equilibrium (Ghil 1983).
The boundary
condition at the scale height 
of the atmosphere (which we take to be
representative of the average elevation where radiation is emitted
from the atmosphere) is then
\begin{equation}
\sigma \frac{\partial \Delta T}{\partial x}|_{x=0}=g\Delta T(x=0)
\end{equation}
We will use the value $g=1.7\ W/m^{2}K$ as used by Ghil (1983) and
Harvey and Schneider (1985).

van Vliet et al. (1980) used Green's functions to solve this model.
The Green's function of the Laplace-transformed
diffusion equation is defined by
\begin{equation}
i\omega G(x,x',i\omega)-\alpha (x)\frac{\partial^{2}G(x,x',i\omega)}
{\partial x^{2}}=\delta (x-x')
\end{equation}
where G is governed by the same boundary conditions as $\Delta T$. 
This equation can be solved by separating $G$ into two parts: $G_{a}$ and
$G_{b}$ with $x < x'$ and $x > x'$, respectively, where $G_{a}$ and
$G_{b}$ satisfy the homogeneous (no forcing) diffusion equation with a
jump condition relating $G_{a}$ and $G_{b}$:
\begin{equation}
\frac{\partial G_{a}}{\partial x}|_{x=x'}-\frac{\partial G_{b}}{\partial x}
|_{x=x'}=\frac{1}{\alpha (x')}
\end{equation}

The power spectrum
of the average temperature in the atmosphere 
in terms of G is given by van Vliet et al. (1980) as:
\begin{equation}
S_{\Delta T_{av}}(f)\propto
Re (\int_{0}^{w_{1}}\int_{0}^{w_{1}} G_{1}(x,x',i\omega)
dx dx')
\end{equation}
\begin{eqnarray}
\propto
Re(\int_{0}^{w_{1}}\int_{0}^{x}G_{1b}(x,x',i\omega)dx dx' \nonumber\\ +
\int_{0}^{w_{1}}\int_{x}^{w_{1}}G_{1a}(x,x',i\omega)dx dx')
\end{eqnarray}
where $G_{1}$ stands for the solution
to the differential equation for G where the source point is located in
the atmosphere. $Re$ denotes the real part of the complex expression.
Two forms of $G_{1a}$ and $G_{1b}$ are necessary for $x$
located above and below $x'$, respectively, due to the discontinuity
in the derivative of $G_{1}$ created by the delta function
(the jump condition).
The solution of $G_{1}$ which satisfies the above differential equation and
boundary conditions is
\begin{eqnarray}
G_{1a}=\frac{L}{\alpha K}(\frac{\sigma 'L}{\sigma L'}\sinh(\frac{w_{1}-x'}{L})
\sinh(\frac{w_{2}}{L'})\nonumber\\ +\cosh(\frac{w_{1}-x'}{L}) 
\cosh(\frac{w_{2}}{L'}))
(\sinh(\frac{x}{L})+\frac{\sigma}{Lg}\cosh(\frac{x}{L}))
\end{eqnarray}
and
\begin{equation}
G_{1b}=G_{1a}+\frac{L}{\alpha}\sinh(\frac{x'-x}{L})
\end{equation}
where
\begin{eqnarray}
K=(\sinh(\frac{w_{1}}{L})+\frac{\sigma}{Lg}\cosh(\frac{w_{1}}{L}))
\frac{\sigma 'L}{\sigma L'}\sinh(\frac{w_{2}}{L'})
\nonumber\\ +(\cosh(\frac{w_{1}}{L})
+\frac{\sigma}{Lg}\sinh(\frac{w_{1}}{L}))\cosh(\frac{w_{2}}{L'})
\end{eqnarray}
and $L=(\frac{\alpha}{i\omega})^{\frac{1}{2}}$ and
$L'=(\frac{\alpha '}{i\omega})^{\frac{1}{2}}$. Performing the integration
van Vliet et al. obtained
\begin{eqnarray}
S_{\Delta T_{av}}(f)\propto
Re (L^{2}(\frac{\sigma 'L}{\sigma L'}
\tanh(\frac{w_{2}}{L})((\frac{gw_{1}}{\sigma}-1)
\nonumber\\
\tanh(\frac{w_{1}}{L})
-\frac{2gL}{\sigma}\frac{\cosh(w_{1}/L)-1}{\cosh(w_{1}/L)}+\frac{w_{2}}{L})
\nonumber\\
+(\frac{gw_{1}}
{\sigma}+(\frac{w_{1}}{L}-\frac{gL}{\sigma}\tanh(\frac{w_{1}}{L}))
((\tanh(\frac{w_{1}}{L})+\frac{\sigma L}{g})
\nonumber\\ \frac{\sigma 'L}{\sigma L'}
\tanh(\frac{w_{2}}{L'})+(1+\frac{\sigma}{Lg}
\tanh(\frac{w_{1}}{L})))^{-1}
\label{bigone}
\end{eqnarray}
For very low frequencies,
\begin{equation}
\tanh(\frac{w_{1}}{L})\approx\frac{w_{1}}{L},
\tanh(\frac{w_{2}}{L'})\approx\frac{w_{2}}{L'}
\end{equation}
\begin{equation}
\frac{\cosh(w_{1}/L)-1}{\cosh(w_{1}/L)})\approx\frac{1}{2}
\frac{w_{1}^{2}}{L^{2}}
\end{equation}
Reducing eq. (\ref{bigone}),
\begin{equation}
S_{\Delta T_{av}}(f)\propto \frac{1}{1+\frac{\omega^{2}}{\omega_{0}^{2}}}
\propto \frac{1}{f^{2}+f_{0}^{2}}
\end{equation}
which is the low-frequency Lorentzian spectrum observed in the Vostok data.
The crossover frequency as a function of the constants chosen
for the model is
\begin{eqnarray}
f_{0}=\frac{g}{2\pi(w_{1}c\rho + w_{2}c'\rho '(1+\frac{g w_{1}}{\sigma}))}
\approx\frac{\sigma}{2\pi w_{1}w_{2}c'\rho '}\nonumber
\\ \approx\frac{1}{25,000\ years}
\end{eqnarray}
which is within an order of magnitude of the observed crossover frequency
of the Vostok data, $f_{0}=\frac{1}{40,000\ years}$.

At higher frequencies
\begin{equation}
\tanh(\frac{w_{1}}{L})\approx\frac{w_{1}}{L},
\tanh(\frac{w_{2}}{L'})\approx 1
\end{equation}
\begin{equation}
\frac{\cosh(w_{1}/L)-1}{cosh(w_{1}/L)}\approx\frac{1}{2}\frac{w_{1}^{2}}{L^{2}}
\end{equation}
then
\begin{equation}
S_{T_{av}}(f)\propto \frac{1}{2}(\frac{2gw_{1}}{\sigma})^{\frac{1}{2}}
(\frac{c\rho\sigma}{c'\rho '\sigma '})^{\frac{1}{2}}
(\frac{g}{w_{1}\rho c f})^{\frac{1}{2}}\propto f^{-\frac{1}{2}}
\end{equation}
as observed.
The high and low-frequency spectra meet at
\begin{equation}
f_{1}=\frac{g}{w_{1}\rho c}(\frac{\sigma}{2gw_{1}})^{\frac{1}{3}}
(\frac{c'\rho'\sigma'}{c\rho\sigma})^{\frac{1}{3}}
4^{\frac{1}{3}}(\frac{c\rho w_{1}}
{c'\rho ' w_{2}})^{\frac{4}{3}}
\end{equation}
\begin{equation}
\approx\frac{1}{10,000\ years}
\end{equation}
which also agrees well with that observed in the Vostok data ($f_{1}\approx
\frac{1}{2000\ years})$.

The time scales of $f_{1}$ and $f_{0}$ correspond to thermal and 
radiative equilibration of the coupled climate system, respectively.
At time scales of 2000 years, the entire atmosphere and ocean are in 
thermal equilibrium. Besides the Vostok data, observational
support for this thermal equilibration time is provided by comparisons
of Antarctic and Greenland ice cores. Bender et al. (1994) found that 
temperature variations in Antarctica and Greenland are correlated above
time scales of 2000 years and uncorrelated below it. This suggests that
2000 years represents the time scale of global thermal equilibrium.

Our model makes a specific prediction of the ability of the oceans to 
absorb an increase in atmospheric temperature resulting from the greenhouse
effect. If the rate of increase of atmospheric temperature produced by 
the greenhouse effect is less than the vertical heat conductivity 
of the oceans, all of the heat produced by the greenhouse effect can be
stored in the oceans for time scales up to 2000 years, the thermal 
equilibration time of the atmosphere and ocean. A 1$^{o}$C per century
warming is seven orders of magnitude smaller than the 
vertical heat conductivity, indicating that the ocean can easily absorb
such an increase. A time scale of 2000 years is much larger than other 
values quoted for the delay in atmospheric warming due to the absorptive
capacity of the oceans. Most studies have suggested time scales on the 
order of decades (Schneider and Thompson 1981). 

We have applied the same model presented in this paper to variations in
the solar luminosity from time scales of minutes to months (Pelletier 1995). 
A stochastic diffusion model of the turbulent heat transfer between the
granulation layer of the sun (modeled as a homogeneous thin layer with a 
radiative boundary condition) and the rest of the convection zone
(modeled as a homogeneous thick layer with thermal and diffusion constants
appropriate to the lower convection zone) predicts the same spectral form
observed in solar irradiance data recorded by the ACRIM project and 
observed in the climate spectra reported here. The time scales of thermal
and radiative equilibrium of the solar convection zone based upon 
thermal and diffusion constants estimated from mixing-length theory 
match those observed in the ACRIM data.

\section{Conclusions}
We have presented evidence that the power spectrum of atmospheric
temperatures exhibits four scaling regions. We presented a model
originally due to van Vliet et al. (1980) 
proposed to study temperature fluctuations in a metallic film
(atmosphere) supported by a substrate (ocean) that matches the observed
frequency dependence of the power spectrum of temperature fluctuations well.
The difference between the high-frequency spectra of continental and
maritime stations may be interpreted in terms of the fact that
air masses above maritime stations 
interact with the atmosphere above and the ocean below  
while air masses above continental stations interact with only the
atmosphere above.

\section{Acknowledgements}
I wish to thank Donald L. Turcotte, Edwin E. Salpeter, and Bruce Malamud
for helpful conversations. I am indebted to Kirk Hazelton, Michael Ghil,
and Pascal Yiou for providing me with data.

\end{document}